\documentclass[letterpaper,english,reprint, aps]{revtex4-2}
\usepackage[T1]{fontenc}
\usepackage[latin9]{inputenc}
\usepackage{amsmath}
\usepackage{amssymb}
\usepackage{graphicx}


\usepackage{babel}
\begin{document}
\preprint{APS/123-QED}
\title{Shape Entropy of a Reconfigurable Ising Surface}
    
\author{Benjamin N. Katz}
\email{bnk120@psu.edu}

\affiliation{Pennsylvania State University}
\author{Lev Krainov}
\email{lxk67@psu.edu}

\affiliation{Pennsylvania State University}
\author{Vincent Crespi}
\email{vhc2@psu.edu}

\affiliation{Pennsylvania State University}
\date{\today}
\begin{abstract}
Disclinations in a 2D sheet create regions of Gaussian curvature whose inversion produces a reconfigurable surface with many distinct metastable shapes, as shown by molecular dynamics of a disclinated graphene monolayer. This material has a near-Gaussian ``density of shapes'' and an effectively antiferromagnetic interaction between adjacent cones. A $\sim$10 nm patch has hundreds of distinct metastable shapes with tunable stability and topography on the size scale of biomolecules. As every conical disclination provides an Ising-like degree of freedom, we call this technique {\it Isigami}. 
\end{abstract}

\maketitle

Techniques arising from papercraft -- origami and kirigami -- inspire efforts to impart three-dimensional shapes to two-dimensional materials \citep{1-Blees2015, 2-Callens2018, 3-Cho2011, 4-Dietz2009, 5-Han2017, 6-Hanakata2016, 7-Morikawa2017, 8-Mortazavi2017, 9-Xu2016, 10-Xu2017, 11-mech_kirigami, Akinwande2017, Wei2018}. Origami and kirigami both work by imposing {\it one}-dimensional modifications (folds or cuts) into surfaces. Here we investigate an alternative method of shape control that inserts {\it zero}-dimensional objects -- isolated disclinations -- into an atomically thin sheet. Inversion of bistable conical disclinations provides a means to control the shape of the surface, with tunable topgraphy down to the length-scale of protein secondary/tertiary structure and energy barriers against shape change that are tunable from below room temperature to values much larger. A tunable topography in a biologically relevant size regime suggests that reconfigurable shape-driven binding interactions may be possible with such sheets.

A disclination removes or adds a wedge of material in the lattice of a two-dimensional sheet, yielding a conical or hyperbolic local region; in a hexagonal lattice the least costly such defects are 5 and 7-membered rings. Since a cone (in a continuum approximation) has a local $C_{\infty}$ symmetry for rotations about its apex, its mechanical states are anticipated to be particularly simple: just ``up'' and ``down'', related by a local inversion. The reduced rotational symmetry of the saddle suggests more complex behavior, as it may be able to assume more than one in-plane orientation depending on the local mechanical environment. Here we investigate a sheet containing multiple disclinations balanced between positive and negative (i.e. asymptotically flat) and determine the multiplicity and character of the shape metastability thereby instilled in the 2D layer. Anticipating our conclusions, the sheet's conformational freedom is dominated by Ising-like bistable cone degrees of freedom; hence we call this means of shape control {\it Isigami}.

\begin{figure*}[t]
\includegraphics[width=6.5in]{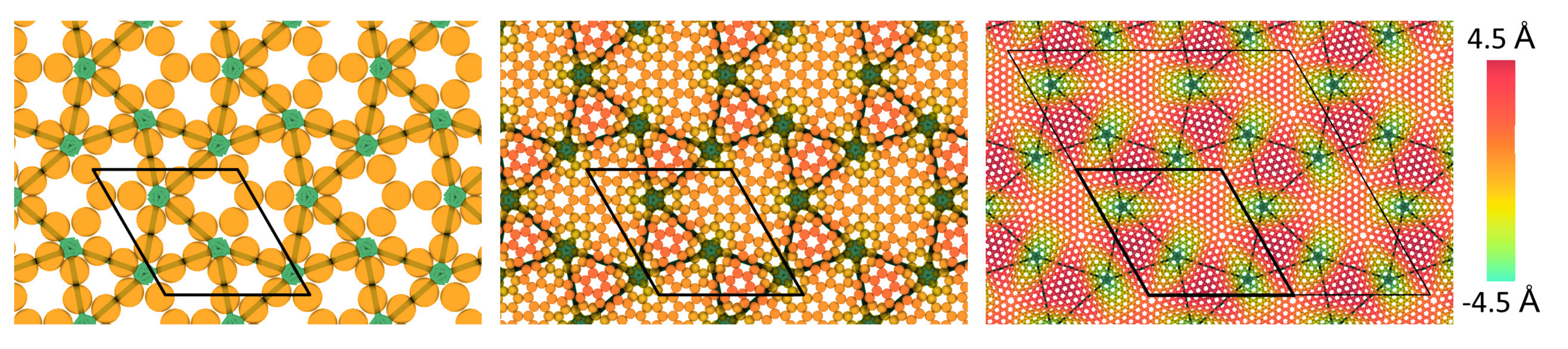}
\label{fig 1}
\caption{Three versions of a graphene-based Isigami sheet with different disclination spacings. The first, with adjoining defects, is a Haeckelite known from \citep{Haeckelite}. The second and third have two- and six-hexagon wide buffers between pairs of disclinations. Sheets are color-coded by height, whose variation grows for wider buffers (the first sheet being perfectly flat). Blue lines delineate a twisted kagome lattice of pentagonal disclinations. The larger unit cell marked on the third sheet shows the size of system most closely studied (in either periodic or finite-patch geometries), containing 12 pentagonal and 12 heptagonal disclinations.}
\label{FigIntro}
\end{figure*}

To create such a surface, we begin with a so-called Haeckelite \citep{Haeckelite,Crespi1996} structure and ``inflate'' each ring with a penumbra of hexagons, similar to inflation of larger fullerenes from $\text{C}_{\text{60}}$ \citep{Dunk2012,Suzuki1991}. We choose a Haeckelite in which the pentagonal rings form a slightly deformed kagome lattice (Fig.~\ref{FigIntro}) \citep{Sun2012}. 
The shape entropy phenomena that we seek to study should not require an ordered lattice of disclinations; this choice is for computational convenience, and to explore whether ideal mechanical frustration may bring the set of possible shapes closer to mutual degeneracy.
Other disclination patterns are also possible, and those formed by phase field crystal methods have been examined independently for their effect on sheet toughness in work by Zhang, Li and Gao \citep{Zhang2014}. We choose graphene as an archetype, since its mechanical response is well-studied and methods exist to functionalize the surface to control mechanical stiffness, interfacial energetics, and hydrophilicity \citep{Singh2016,Wu2015,Morimoto2016}.  Since the behavior described below is largely geometrical in origin, conclusions derived from graphene Isigami should generalize to other atomically thin two-dimensional materials.

We construct both periodic and finite-patch regions of material hosting twelve cones and twelve saddles on a kagome lattice as depicted in Fig.~\ref{FigIntro} and model the mechanical response using the AIREBO potential \citep{AIREBO} as implemented in LAMMPS (Large-scale Atomic/Molecular Massively Parallel Simulator)~\citep{LAMMPS}, as it is well-validated for the mechanical response of nanoscale sp$^2$ carbon~\citep{Svatek2014, Gu2013, Meguid2017, Giordanelli2016, Nanochimneys}. We first examine a finite sheet whose disclinations are separated by six hexagons ($\sim$20 \AA); the edges are terminated by hydrogen. This separation is large enough that even cones on the edge of the sheet have two well-defined metastable ``up'' and ``down''  orientations. Shape variation is most dramatic in a finite patch of material, as the periodic boundary constraint is absent. The patch is systematically forced into every possible set of cone orientations by mechanically inverting various sets of cones. To this end, a force of $\pm1.7$ eV/\AA\ is applied to the five atoms at the apices of various cones according to the desired up/down states and the system is then allowed to evolve for 500 femtoseconds while the atoms are constrained to move no more than $0.1$\AA\ each timestep. A constraint force is also applied evenly across every atom in the system to zero-out the net force. The structure is then relaxed either through a thermal bath at 300 K for 50 ps followed by a linear ramp anneal over 30 ps from 300 K to 3 K, or (for the periodic systems described below) through a shorter 1 ps thermal bath followed by a series of conjugate gradient minimizations where the unit cell geometry is allowed to relax during alternate optimization steps. These methods both produce well-converged energy minima (converged to less than 100 meV in systems with thousands of atoms), but the anneal method performs better for the finite patch in avoiding local minima.

\begin{figure}[h]
\includegraphics[width=3in]{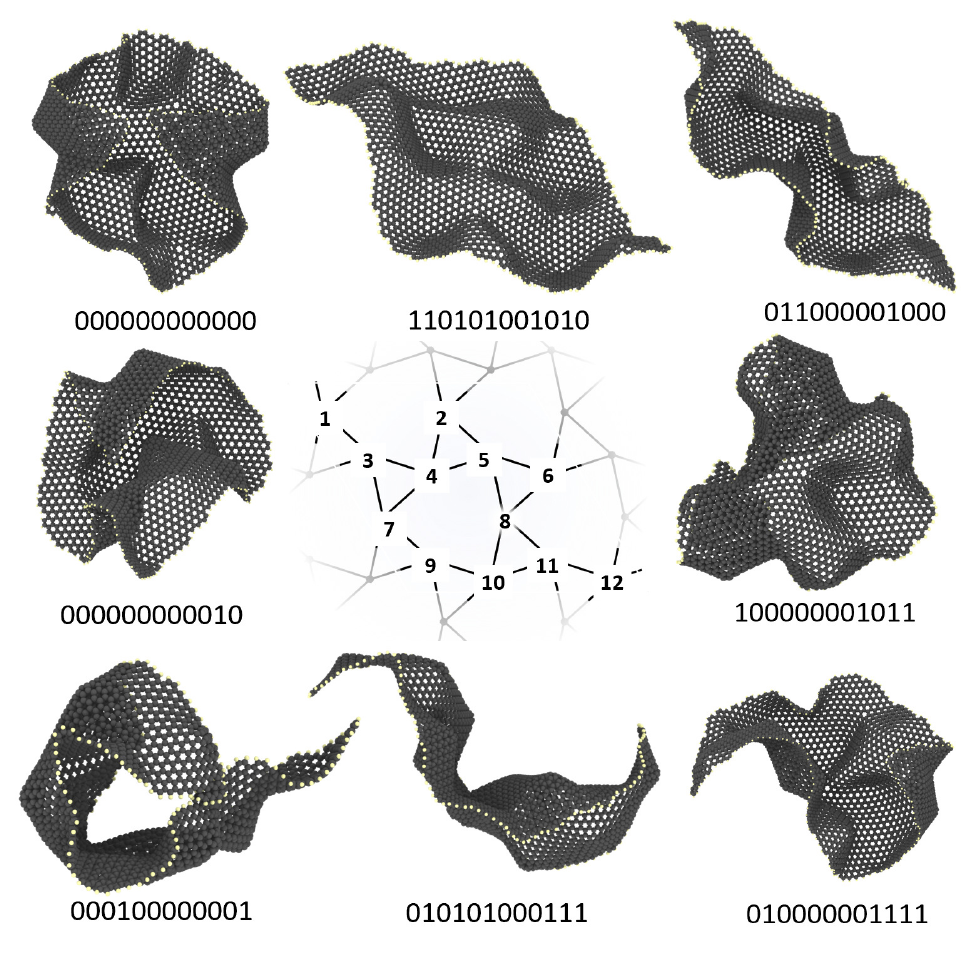}
\label{fig 2}
\caption{Representative shapes for a 12-cone patch, as a demonstration of the multiplicity of metastable shapes. The shapes are displayed in their minimum energy states, and are hydrogen-terminated (yellow atoms) with carbon atoms in gray. A bitstring of cone states in the sheet is displayed below each example; this labeling follows the order shown at center.}
\label{mult}
\end{figure}

For every initial choice of the twelve up/down cone configurations, the system relaxes into a distinct metastable shape, i.e. all $2^{12}$ nominal Isigami configurations are accessible. Fig.~\ref{mult} shows some examples of these shapes, labeled by the cone configuration: the different shapes that are metastably assumed by the same sheet can be dramatically different. Several of these configurations, including the five highest in energy, were simulated at 800 K in an NPT ensemble for one nanosecond, with the sheet maintaining its original Ising state. This high-temperature ``challenge'' anneal was also applied to a subset of configurations that were repeated to create a $2\times 2$ supercell patch, providing assurance that shape metastability is preserved for larger patch sizes as well. 

Collapsing the reflection symmetry about the sheet midplane, we actually obtain $2^{11}$ distinct Isigami configurations for this 12-pentagon patch, with an additional approximate reflection symmetry about the short diagonal (broken very weakly by the detailed termination of the patch). Additional symmetries in the periodic case further reduce the number of symmetry-distinct configurations per unit area, but the conformational entropy of metastable Isigami states remains extensive.

\begin{figure}[th]
\includegraphics[width=3.5in]{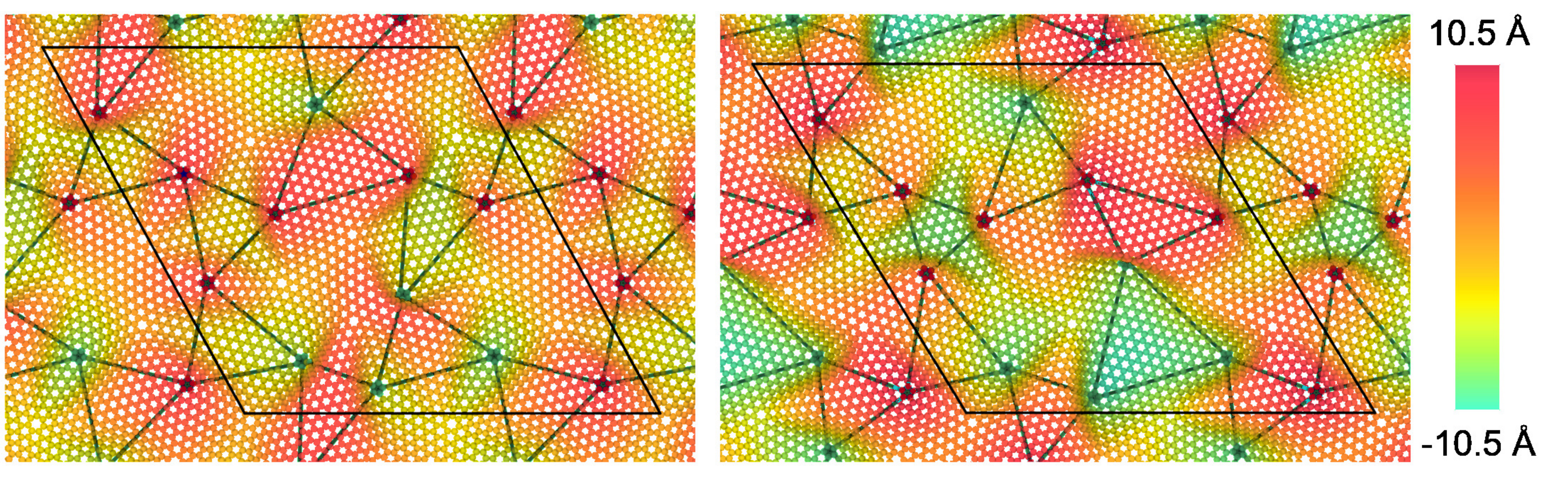}
\label{fig 4}
\caption{Two periodic states with the same cone orientations but different metastable shapes: the state on the right is 6.5 eV lower in energy. The images are color-coded by height on the same scale, and the deformed kagome lattice of each is shown in blue. Cones pointing up have red apices; those pointing down are blue.}
\label{motif}
\end{figure}

As noted earlier, whereas cones have two distinct conformations (i.e. up and down), the metastable orientations of the principal axes of saddle-point disclinations are not naturally two-fold; their number depends on their surroundings. We do in fact find certain conformations that have the same set of cone orientations (i.e. the same Isigami state) but visibly different overall shapes, sometimes differing by over an eV in energy and stable in shape up to at least 800 K over half a nanosecond. Fig.~\ref{motif} depicts two such states. While we have not exhaustively quantified the frequency of these non-Ising states, our observations suggest they contribute a number of configurations equivalent to roughly one additional Ising degree of freedom for a 12-cone sheet; i.e. their contribution to the overall configurational entropy is modest. Although these non-Ising states do sometimes correspond to different orientations of the saddle disclinations, we do not find a simple rule governing their appearance, as reflected by the examples shown in Figure S5. Similar non-Ising states are seen in the larger $2\times 2$ supercell patch. Simulations of this larger patch (and larger periodic systems) reveal a similarly pervasive metastability to Ising conformations, suggesting that the system has an extensive conformational shape entropy. 

\begin{figure}[h]
\includegraphics[width=3.5in]{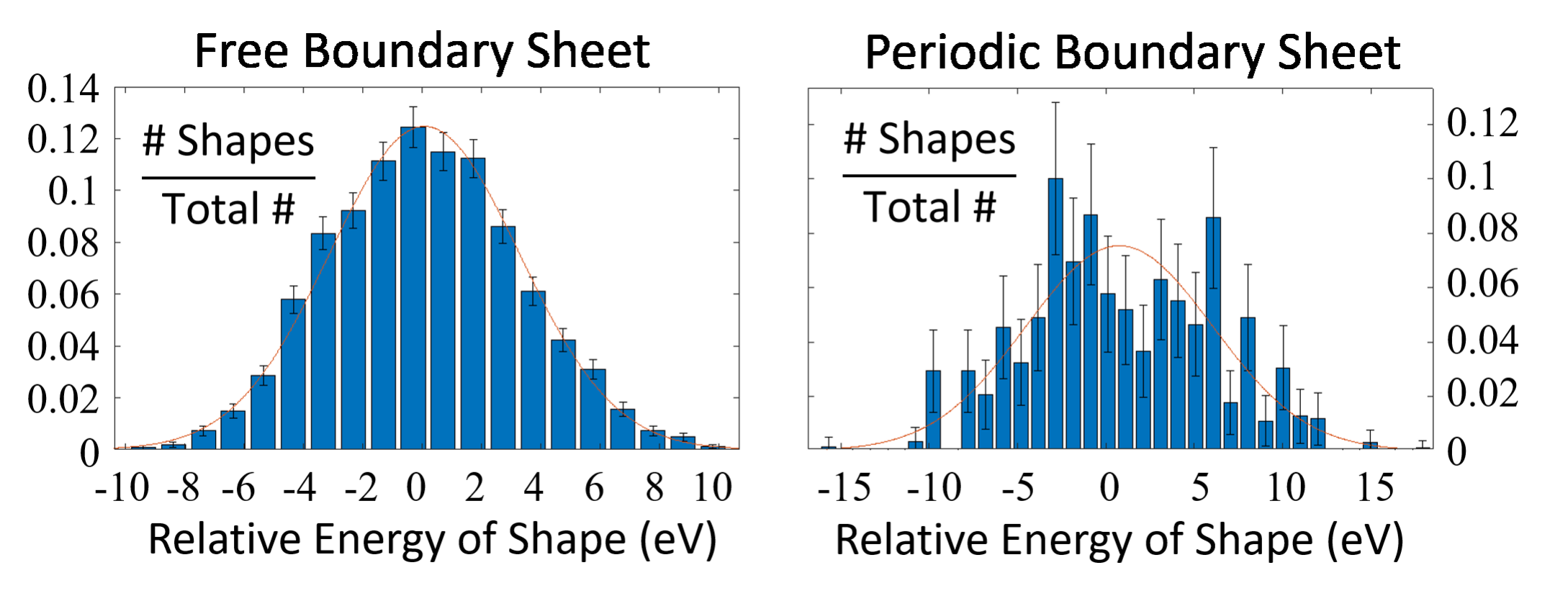}
\label{fig 5}
\vspace*{-0.25in}\caption{The number of shapes of the 12-cone patches studied versus their relaxed energy, as a fraction of the total number of shapes examined for both finite and periodic sheets, with a normal distribution superimposed. The error bars follow $\sqrt{N}$.}
\label{gaussian}
\end{figure}

How are these states distributed in energy? The ``density of shapes'' (i.e. the number of shapes in a given energy range) for the finite 12-cone patch is strikingly close to Gaussian, as shown in Fig.~\ref{gaussian}. It is tempting to consider this outcome as an expression of the central limit theorem for multiple nearly independent energetic contributions from different disclinations, but the stronger deviations from a Gaussian shape seen in a similarly-sized periodic patch argue against a simple application of this notion. The distribution for the finite patch may be more Gaussian because its boundary condition produces multiple symmetry-distinct pentagonal disclinations within the patch and these heterogeneous local environments, under the action of elastic interactions that span the entire patch, yield something closer to the requirements of the central limit theorem. In contrast, every pentagonal disclination under periodic boundary conditions is symmetry-equivalent. For less well-ordered disclination networks we anticipate the periodic case to more closely approach the finite-patch distribution.

\begin{figure}[h]
\includegraphics[width=3in]{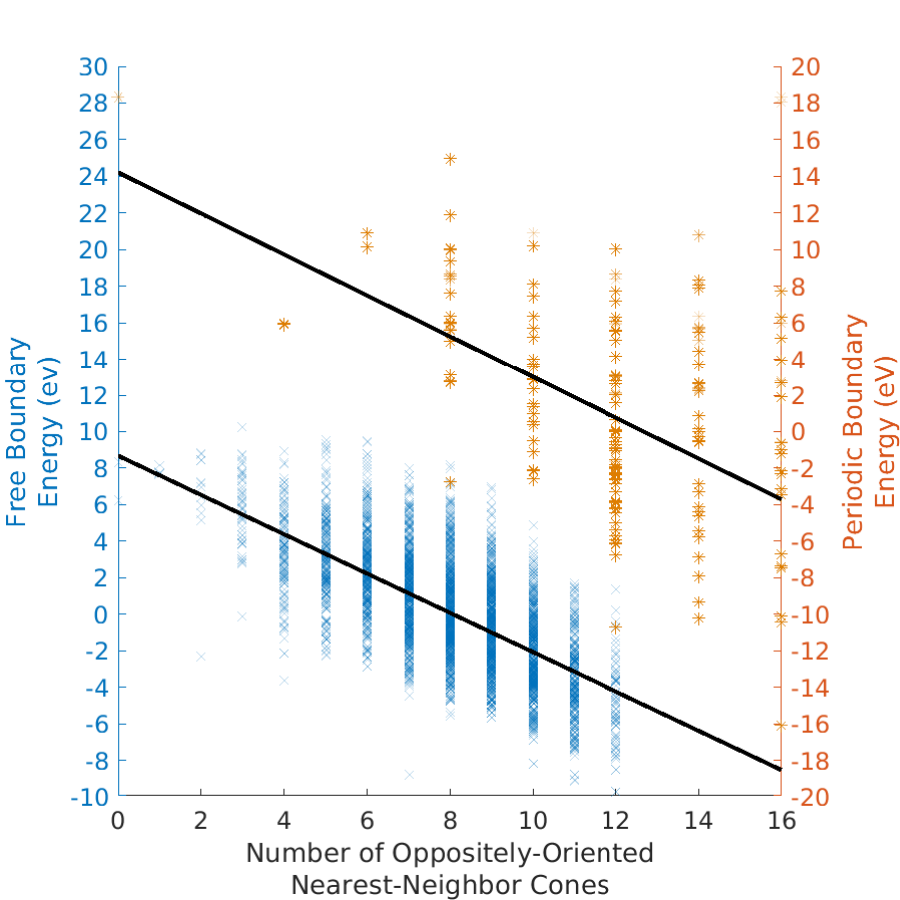}
\label{fig 3}
\caption{The sheet energy as a function of the number of nearest-neighbor cone pairs which are oppositely-oriented for both free (left, blue) and periodic (right, orange) boundary conditions, with least-squares linear fits.  A nearest-neighbor pairwise antiferromagnetic interaction between pentagonal disclinations would yield a linear dependence, buried under significant variation from longer-ranged interactions, boundary effects and a modest admixture of non-Ising states, as seen. The linear trend yields similar nearest-neighbor antiferromagnetic $J$'s for free ($-0.54$ eV) and periodic ($-0.56$ eV) cases. The free-boundary case has fewer degeneracies and hence more distinct points. Points are translucent to better reveal the overall density of configurations.}
\label{antiferro}
\end{figure}

Mechanical intuition suggests that nearby cones will have pair-wise antiferromagnetic interactions, since opposing up/down orientations produce compatible sidewall slopes. In practice, we find that longer-range and many-body elastic interactions between disclinations are too strong to admit a strict frustration-derived degeneracy. While these non-idealities (and also non-Isigami states) are sufficiently important that the sheet is not a simple nearest-neighbor antiferromagnet, the energies of different sheet configurations {\it do} reveal an overall nearest-neighbor antiferromagnetic trend, as the energy varies systematically with the number of oppositely-oriented nearest-neighbor cones for both the periodic-boundary and finite-patch cases (Fig. \ref{antiferro}). The effective $J$ of this trend is proportionately smaller for a sheet with 1-hexagon separation between the 5-fold rings (Fig. S1). As noted in Supplementary Information, it was not possible to describe the sheet energetics with a simple cluster expansion~\citep{Sanchez1984}; this was likely due in part to the long-range nature of elastic interactions amongst nearby cones (i.e. when even a single cone inverts, we see significant changes in shape across the entire patch, as shown in Figure S4) and in part to the presence of a modest number of additional states not fully specifiable by an Isigami state vector, as noted earlier. 

\begin{figure}[h]
\vspace*{-0.1in}\includegraphics[width=3in]{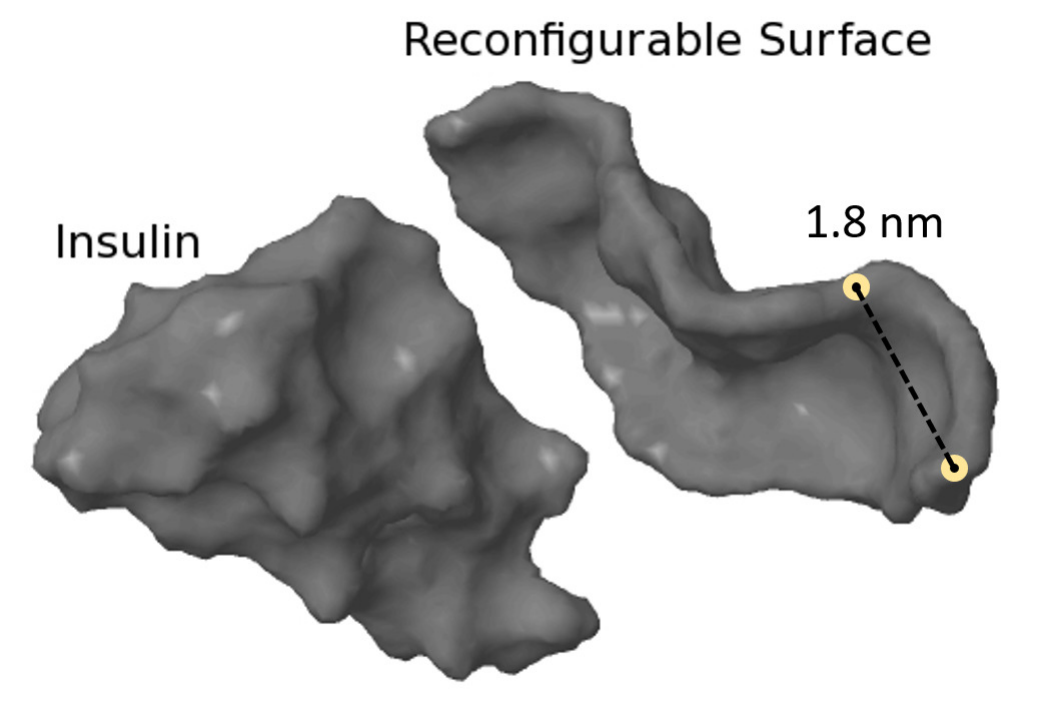}
\label{fig 6}
\vspace*{-0.1in}\caption{Comparison between an Isigami sheet (at right) with a two-hexagon buffer between cones and a generic small protein (insulin, at left) that has been smoothed using the Connolly (``rolling ball'') method for constructing solvent-excluded surfaces, done here with a solvent radius similar to the bending radius of the sheet to accentuate the topography of both across the same length scale, showing it is possible to construct Isigami sheets with reconfigurable topography on the same length scale as protein secondary/tertiary structure.}
\label{sheet-protein comparison}
\end{figure}

These reconfigurable surfaces provide a means to access a multitude of distinct sheet topographies from a single patch of multiply disclinated 2D material: an obvious area of potential application for such an object is shape recognition and binding of biomolecules. Ideally, such a surface would have topography on the length scale of protein tertiary structure, a hydrophilic surface that does not denature biomolecules, and a tunable barrier against shape change so that both dynamic ($\sim k_BT$) and static ($\gg k_BT$) regimes can be accessed. The $\sim 0.5$ nm inter-pentagon spacing of the sheet depicted in Fig.~\ref{sheet-protein comparison} provides topography on a biologically relevant length-scale, as demonstrated by the comparison to insulin, a small protein, in the same figure. The use of graphene {\it oxide} \citep{Pandit2020,Kenry2016} rather than hydrophobic bare graphene~\citep{Munz2015, Alava2013, Feng2019} could introduce the needed hydrophilicity, and perhaps also increased flexibility~\citep{Poulin2016}. 

The question of barrier height control requires more detailed study. The inversion barriers of isolated graphene cones have been simulated previously~\citep{cone_inversion}; the barrier depends on the chemical character of the conical apex but is generally multiple electron volts for well-separated cones. Energy barriers against cone inversion will depend on the material of the sheet, the separation between conical disclinations, functionalization~\citep{cone_inversion}, or grain boundaries interconnecting disclinations~\citep{Liu2010, Wang2017}. We observe in the periodic sheets that the stability of the Ising and non-Ising states indeed varies with cone separation, and also that the barriers against inter-conversion of the non-Ising states are typically much lower than those between Ising configurations. We monitored the energy during cone inversion for a few periodic sheets with different cone separations; although the distribution of inversion barriers could not be exhaustively quantified, the barriers all remained on the order of electron volts, likely lowered somewhat (compared to isolated cones) by irregularities in the local environment. The typical barrier height scales roughly linearly with the separation between pentagonal disclinations, being around 1 eV for a 1-hexagon separation and around 6 eV for the 6-hexagon separation. At the 1-hexagon separation a significant fraction of the Isigami states destabilize (see Figure S2) while almost all of the non-Ising states disappear, where our criterion for stability is surviving the room-temperature anneal mentioned earlier. 

Since most of the cases considered above have inversion barriers much greater than $k_BT$, we now consider a means to ``thermalize'' these barriers even at larger disclination separations: oxidizing the sheet and etching away the central five-membered ring and one or more surrounding rings to reduce the mechanical deformation associated with nucleating inversion, terminating the resulting apical multivacancy with hydrogen atoms. To this end, we oxidized a six-hexagon-separation periodic sheet by random placement of oxygen atoms equally onto upper and lower surfaces followed by a structural relaxation and subsequent removal of any unreacted oxygen, yielding a functionalized sheet with 12.6 at\% oxygen (Fig. S3).  The apices of this sheet up to three surrounding rings were etched away by manual deletion, and the resulting edges terminated with hydrogen.
These simulations used a version of the REAXFF reactive force field \citep{Duin2001, Velde2001, ADF2019} designed \citep{Chenoweth2008} to describe carbon, hydrogen and oxygen \citep{Yang2020} and updated to more precisely describe chemical properties of C/H/O/N systems \citep{Kowalik2019, Srinivasan2015, Ashraf2017, Privcomm}. Neither the structural relaxation following etching (nor that which followed oxygenation) caused cone inversion, but during a subsequent MD simulation at 800K we observe rapid ($\sim 1$ ns) spontaneous inversion of cones; we expect a similar simulation without periodic boundary conditions to be at least as floppy. Although these simulations were performed in the gas phase, we anticipate that the qualitative conclusion of low, thermally accessible inversion barriers will be robust against the introduction of aqueous conditions (which may themselves further reduce the barrier). In constrast, the same system with un-etched apices shows no inversion over 1 ns at 800 K. 

Shape reconfigurability at the nanometer-to-micron scale is a signature of biology, but one that is not straightforwardly expressed in non-biological systems. We describe a strategy towards this end that uses the bistability of disclinations to afford a large library of shapes from a single 2D sheet. Selection from this library could either happen spontaneously -- for example, through binding interactions with targets under conditions of shape lability -- or be imposed through mechanical deformation by a specified indenter. We demonstrate the second of these through large-scale molecular dynamics simulations of indented Isigami sheets. The simulated sheets are large enough to be compatible with experimental nano/micromechanical manipulation~\citep{liu2010a,sheng1999} (i.e. hundreds of nanometers across) and make use of simple radial and annular indenter geometries, as depicted in Supplementary Videos S1-S3. The three sheets simulated all start from the same asymptotically flat Ising configuration.
The indenters' pressures force all the cones underneath to orient similarly, which creates linear structures of unbalanced (i.e ferromagnetically aligned) Ising state. The antiferromagnetic interaction between neighboring cones in the indented region then drives a sheet distortion that splays these cones away from each other to minimize their unfavorable co-alignment: the resulting crumples ``bunch up'' material under the indented region. For the radial indenter this bunching reduces the circumference at fixed radius from the center of the indenter, while the annular indenter does the opposite; thus they create large-scale Gaussian curvatures of opposite sign. The cross-shaped identers of different widths also produce different bend angles. We apply four-fold symmetric indenters to show that vertices defined by the intersection of four folds can be easily created, while a generic crumpled fold would involve only three. These deformations should be reversible by compressing between flat planes and re-indenting.

As this approach is purely geometrical, it should apply to any atomically thin sheet, which could be polar (hBN), hydrophilic (graphene oxide \citep{Pandit2020, Kenry2016}) or stiffer than graphene \citep{Lai2016}. 
These phenomena should also occur in more irregular disclination networks that could be produced by e.g. growing a 2D layer on a roughened substrate, as strict degeneracies due to mechanical frustration on a regular lattice were neither observed nor required: mechanical bistability is a local property of each conical disclination, one whose antiferromagnetic near-neighbor interactions then favor a large shape multiplicity. The degree of metastability is tunable from highly stable to potentially room-temperature reconfigurable. Possible longer-term applications for 2D shape reconfiguration could involve diverse shape libraries exposed to binding targets under shape-labile conditions, selected for binding, and then utilized under shape-conserving conditions. On longer length scales, linear arrays of disclinations could recover phenomena akin to more traditional fold-based origami wherein mean curvature plays a more prominent role. The presence of disclinations (with associated geometrical phases \citep{Lammert2004}) and local strains (with associated pseudomagnetic fields \citep{Levy2010, Gill2015, Zhang2018, Zhang2019}) may also introduce unusual electronic and plasmonic \citep{Kang2018} properties. 

\begin{acknowledgments}
B.N.K. acknowledges training provided by the Computational Materials Education and Training (CoMET) NSF Research Traineeship (grant number DGE-1449785). B.N.K., L.K. and V.H.C. acknowledge support under NSF awards DMR-2011839 and DMR-2039351. Many thanks to Dr. Malgorzata Kowalik for her assistance with ADF and ReaxFF potentials in general.
\end{acknowledgments}

\appendix

\bibliographystyle{apsrev4-2}
\bibliography{shape_memory}

\end{document}


\begin{center}
{\Large \bf Supplementary Material}
\end{center}

\vspace{3mm}
\noindent {\bf \large Analysis of barrier height for cone inversion: }
Approximate magnitudes for transition barriers were calculated within LAMMPS using the AIREBO potential. A few selected structures covering a variety of local curvatures and total energies were chosen from the periodic lattice as a representative sampling. For each of these, one cone at a time was slowly moved from its original position to an inverted one. Two reference atoms in the sheet, one at the cone apex and one elsewhere in the sheet (and well-separated from the cone in question) were fixed in the direction perpendicular to the sheet plane to set vertical height to act as a reaction coordinate and explore minimally constrained inversions. Different choices for the fixed atom in the sheet produced modest changes in overall barrier height. The typical inversion barrier is around 6 electron volts for the sheet with six hexagons of separation between apical disclinations and around 1 eV for the one-hexagon case. A more systematic sampling of barrier heights was complicated by the frequent inversion of a cone other than that specified by the apical fixed atom.

\begin{figure*}[t] 
\includegraphics[width=5in]{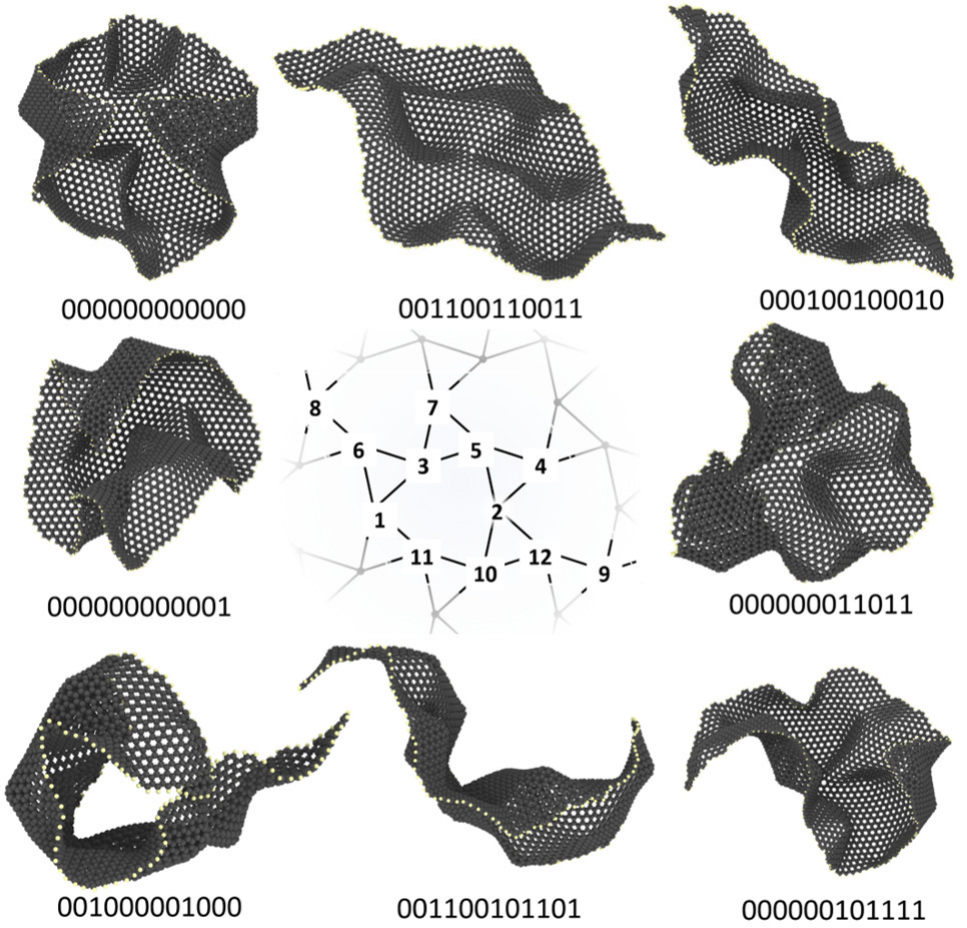}
\caption{Scripts provided elsewhere in Supplementary Information make use of the cone numbering scheme shown above, which differs from the streamlined version used in Fig. 3 of the main text.}
\label{configs}
\end{figure*}

\begin{figure*}[t] 
\includegraphics[width=6.5in]{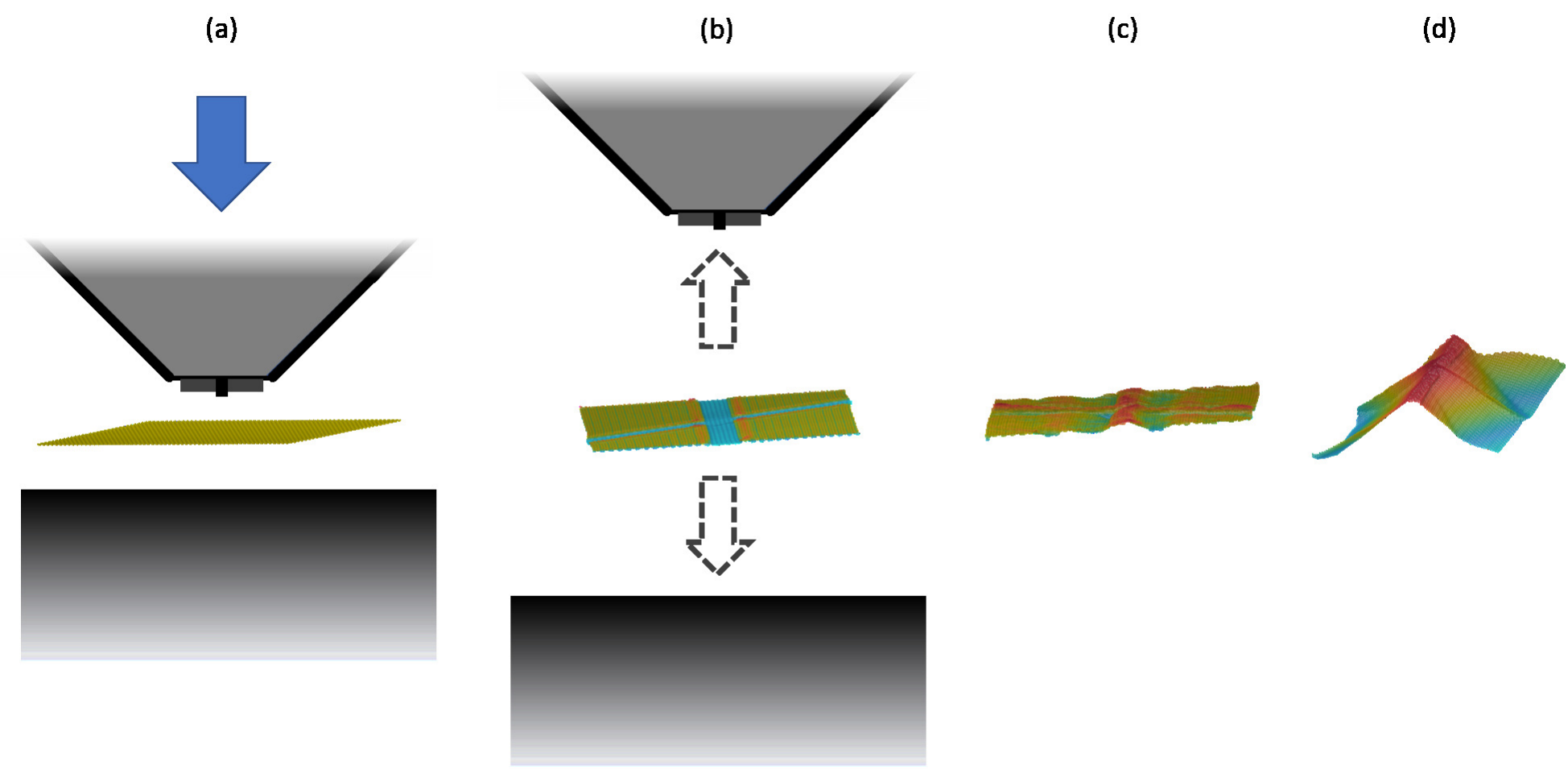}
\caption{A schematic of the MD simulation, in which: (a) an indenter descends, pressing an initially asymptotically flat sheet into a Van der Waals substrate, (b) the indenter and substrate are removed, and (c) the sheet is left to evolve under NVT dynamics at 300K until (d) it reaches a final relaxed state.}
\label{schematic}
\end{figure*}

\begin{figure*}[t] 
\hspace*{-0.25in}
\includegraphics[width=7in]{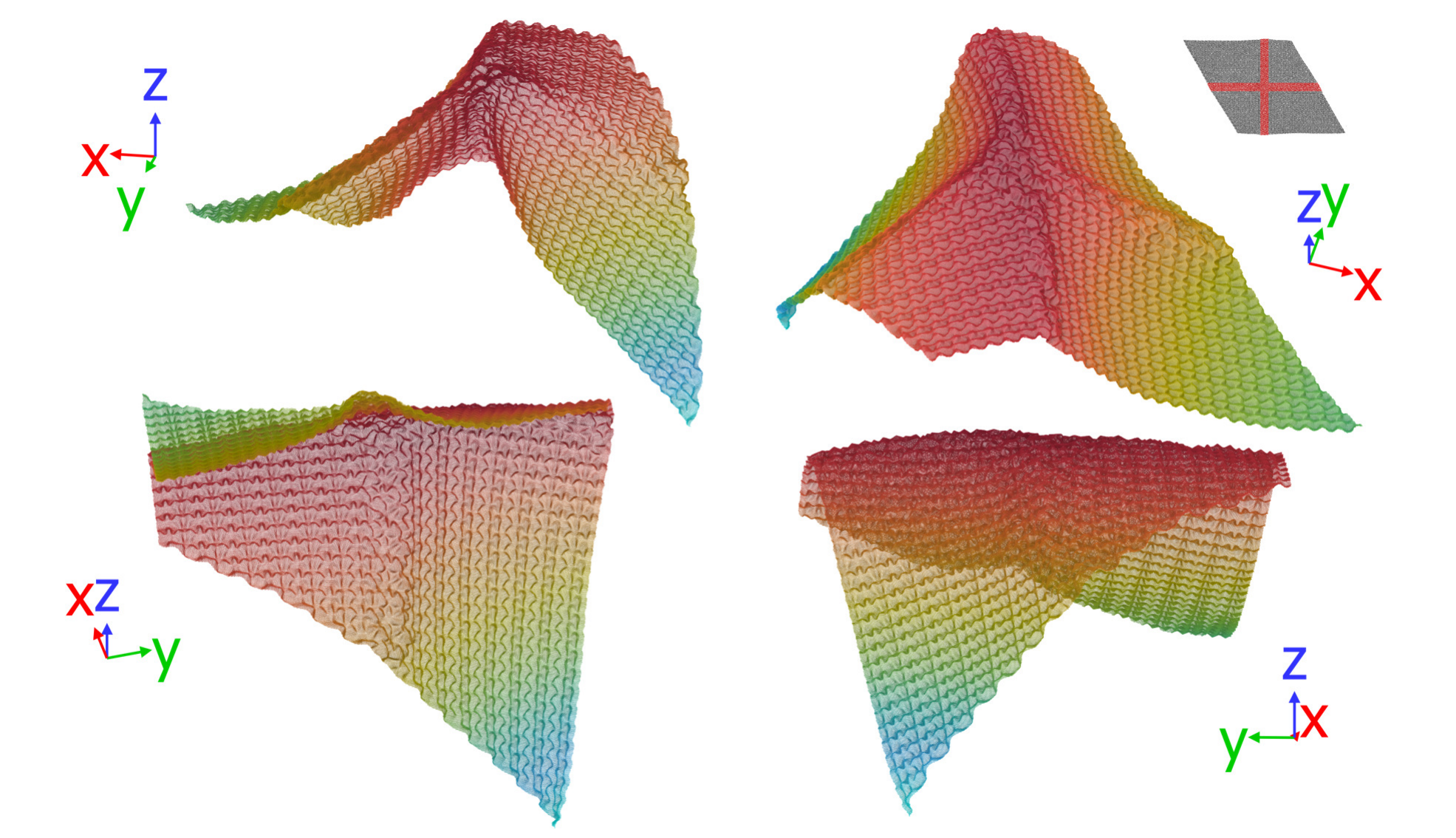}
\caption{Views of the surface evolved in video S1, a shape with long-range positive gaussian curvature within a rhombus-shaped sheet with side lengths $\sim$127.5 nm. The deformation is formed by a radial indenter with stripes of 10 nm width (the indented area is highlighted in the inset). Bend angles produced by the folds, roughly defined, are marked.}
\label{cone_1x}
\end{figure*}

\begin{figure*}[t] 
\hspace*{-0.25in}
\includegraphics[width=7in]{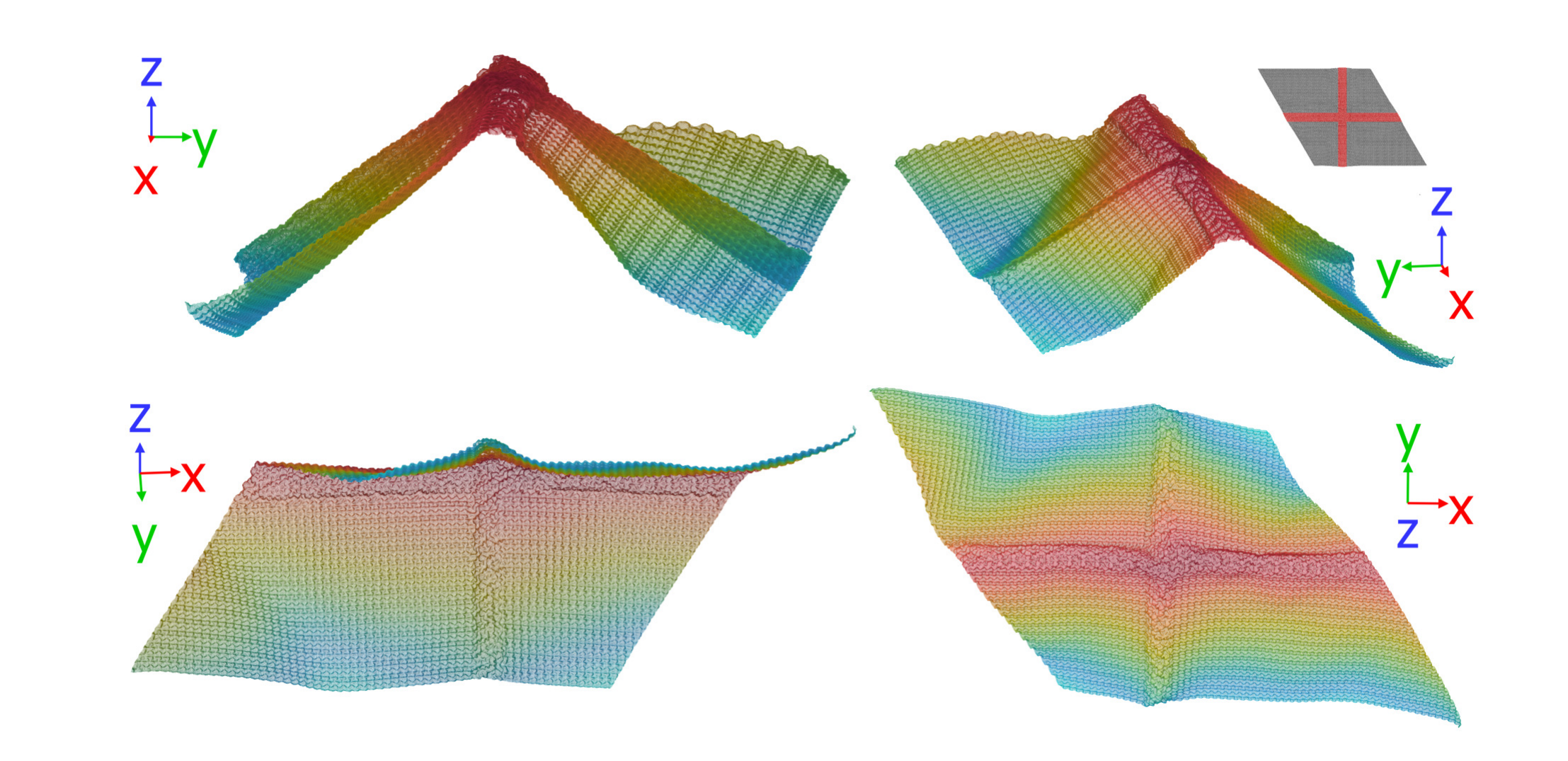}
\caption{Views of the surface evolved in video S2, a shape with long-range positive gaussian curvature made from rhombus a with side lengths $\sim$253 nm, formed with a radial indenter with stripes of 20 nm width (the indented area is highlighted in the inset). Note the larger bend angle as compared to Fig. \ref{cone_1x}, even though the relative area of indentation to unmodified sheet is the same: the bend angle depends primarily on the width of the indented strip.}
\label{cone_2x}
\end{figure*}

\begin{figure*}[t] 
\hspace*{-0.25in}
\includegraphics[width=7in]{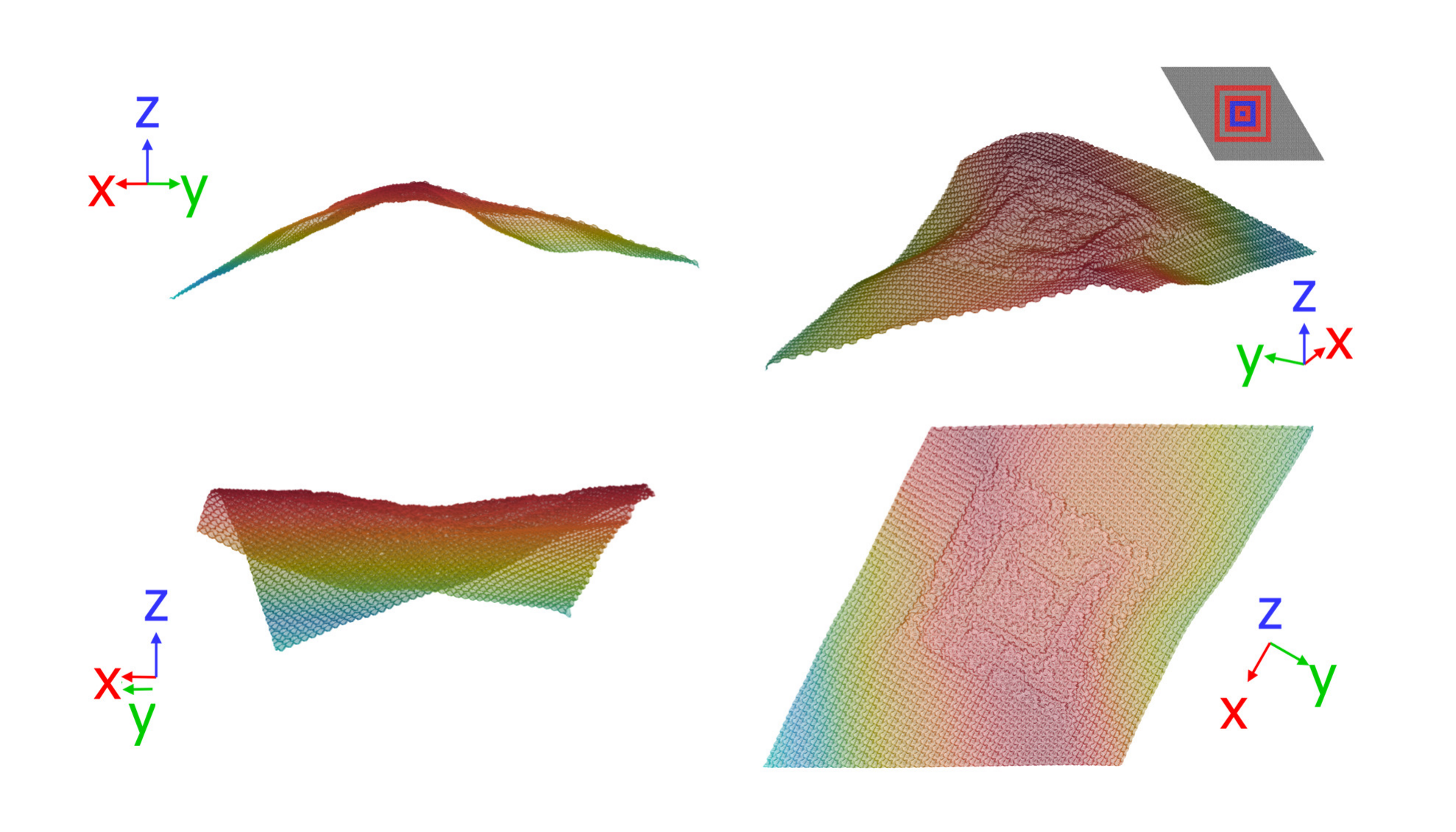}
\caption{Views of the surface evolved in video S3, a shape with long-range negative gaussian curvature made from a rhombus with side lengths $\sim$211 nm, formed with an annular indenter comprising concentric squares of 10 nm width (the indented area is highlighted in the inset).}
\label{saddle}
\end{figure*}

\begin{figure*}[t] 
\includegraphics[width=5in]{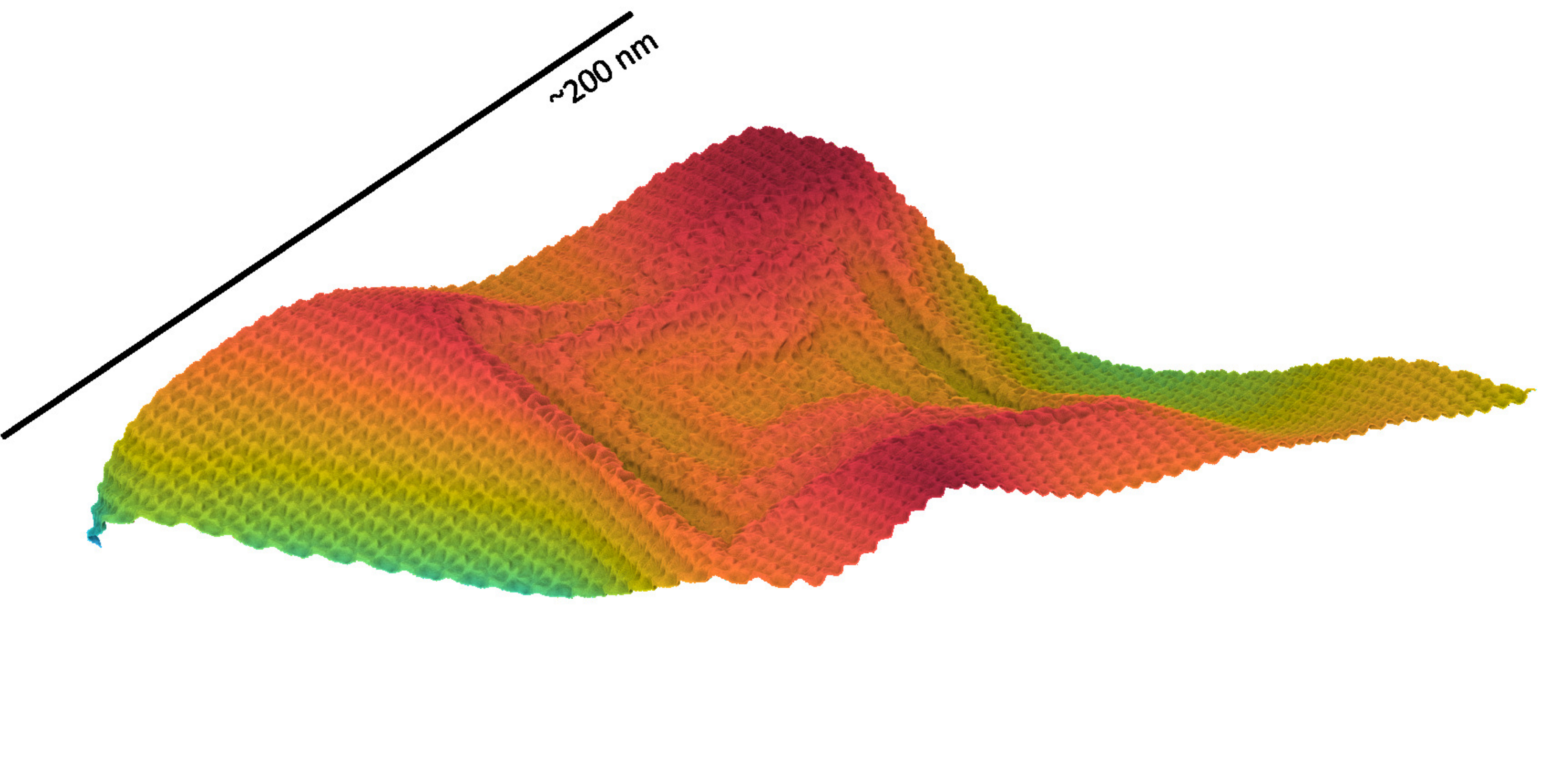}
\caption{Image of a saddle formed by selected concentric inversion of cones in the surface, with a scale bar for size.}
\label{old_saddle}
\end{figure*}

\begin{figure*}[t] 
\includegraphics[width=5in]{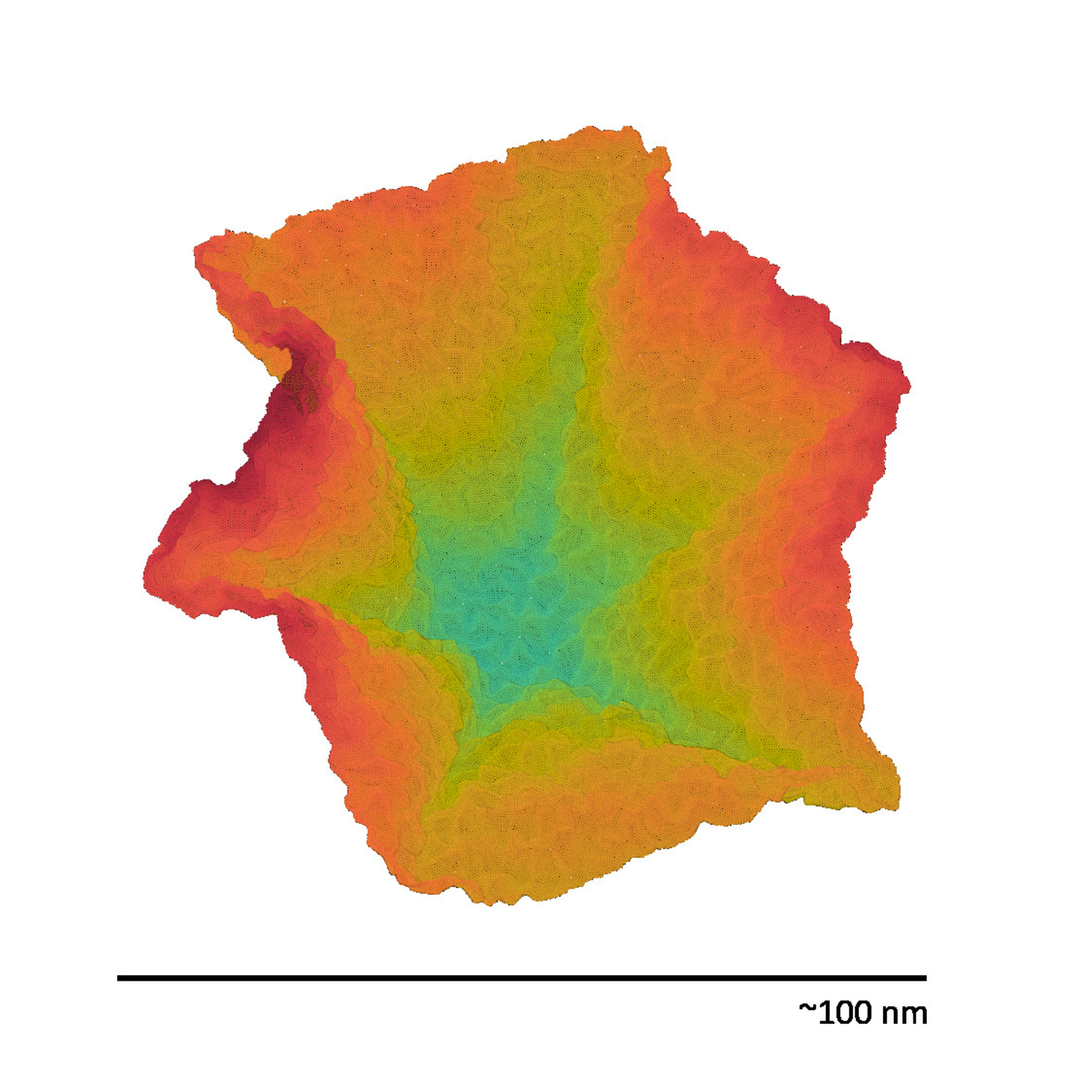}
\caption{Image of a bowl-like shape formed by selected inversion of cones in the surface of a circular patch of originally flat material, with a scale bar for size.}
\label{old_cone}
\end{figure*}

\vspace{5mm}
\noindent {\bf \large Description of video simulations:}
Included in online supplementary material are three videos showing large Isigami sheets (i.e. of a size  accessible to micromechanical manipulations) whose shapes can be controlled through mechanical manipulation by micron-scale indenters.  The videos show three sheets starting from the same asymptotically flat Ising configuration: the top-center configuration shown in Fig. \ref{configs}.
Snapshots from the three videos are given in Figures \ref{cone_1x}, \ref{cone_2x}, and \ref{saddle}, with a schematic of the simulated indentation in Figure \ref{schematic}.  Two images from more ad hoc  simulations (Figures \ref{old_cone} and \ref{old_saddle}) are also shown; these were created with an eye to particular endpoints.

As discussed in the main text, neighboring cones interact antiferromagnetically.  If several rows of cones arrayed in a stripe are all oriented identically, they will then have a repulsive interaction with their neighbors.  This causes the sheet to bend along the line of the stripe to allow the cones to splay outward and thereby point away from each other to a greater degree, as would antiferromagnetically interacting magnetic dipoles oriented normal to a curved sheet.  Mechanically, this is reflected in the cones transitioning from an oval to a more circular cross-section as the stripe relaxes into a fold. 

This bending then induces gaussian curvature on the scale of the whole sheet, because each of several radial folds effectively reduces the circumference measured around the central point at some fixed radius (think of how the folds in a tablecloth draped over a circular table would extend up to the apex if draped over a conical table). We can do the opposite with annular folds, as these folds effectively transport sheet material inwards towards the center, thus yielding too-much material at a given radius from the center, material whose circumference can only be fit by deforming the sheet into a saddle shape.  The outer corners of each sheet may then reactively curl oppositely, depending on the detailed shape of the sheet (see particularly Figure \ref{old_cone}).

We simulate both radial and annular indenters, placing the two radial identers (of different widths) each above a flat substrate and the annular indenter above a substrate with complementary interlocking annular shapes. Indentation is simulated by applied forces on atoms in a specified region while the entire sheet is supported from behind by a Van der Waals barrier.  In all three cases, the indenter presses on the sheet for two hundredths of a nanosecond with a pressure of 1.6 GPa (lower than maximum pressures seen in AFM tips \citep{liu2010a}) before being removed, at which point the substrate is also removed and the sheet is allowed to anneal at 300 K in an NVT ensemble for approximately 2 nanoseconds for both radial indenters and 2.6 nanoseconds for the annular indenter. These simulations were performed in LAMMPS \citep{LAMMPS} using the AIREBO potential \citep{AIREBO}. Indentation forces all the cones under the indenter to orient in the same direction.  This change in Ising cone configuration then causes the sheet to bend around the indented stripes: the sheet thus changes shape because of the change in the Ising state, and this shape change is persistent. The simulated indenters are intended to be in the size range of AFM tips \citep{Sheng1999} or other micromechanical manipulators. As the inversion barrier is roughly linear in the cone radius yet the applied force is proportional to radius squared (for constant indenter pressure), the process of cone inversion is anticipated to become easier for more widely spaced cones: the sheet here, although physically larger than those discused in the main text, uses the same 6-hexagon cone spacing. The two sheets with radial indenters have a single cross-shaped indenter pressing down onto a sheet shaped like a rhombus with side lengths of either 127.5 nm (for an indenter with a 10 nm width to each arm of the cross) or 253 nm (for an indenter of 20 nm width).  The third sheet, a rhombus with side 211 nm side length, is pressed by two interlocking square annular indenters each of 10 nm width with a secondary Van der Waals backplane to simulate the inside surface of the indenter. The second-outermost ring of this annular indenter does not have an upwards component, only the Van der Waals planes: as can be seen during the simulation, the sheet correspondingly folds less strongly around this area than it does around the others, though the fold still persists.

The sheets pressed by the two cross-shaped indenters fold similarly, but with bends of different angles due in large part to the different in crossbar width: wider crossbars produce sharper folds.  While the crossed indenters produce overall positive Gaussian curvature in the folded sheet, the square-annular indenter produces a saddle shape of overall negative Gaussian curvature at long length-scales, as anticipated. Various combinations of these geometries and others provide a basis for constructing alternative shapes such as the saddle and ``taco bowl'' shapes depicted in Figures \ref{old_saddle} and \ref{old_cone}. These two shapes were formed with controlled cone inversion without explicit indenters, to show some of the diversity of metastable shapes attainable by Isigami.

\vspace{5mm}
\noindent {\bf \large Attempted cluster model:}
The cones' twisted kagome lattice has significant geometrical frustration under antiferromagnetic nearest-neighbor spin interactions, which motivates a study of whether our mechanical system can be similarly described, i.e. whether their energetics is dominated by nearest-neighbor pair-wise antiferromagnetic mechanical interactions, and whether the overall shape of the membrane can be described solely with Ising-like degrees of freedom for the conical disclinations. To assess this, we attempted to fit the energies of the periodic system with a Hamiltonian using cone clusters of various sizes~\citep{Sanchez1984}:
\[
H=\sum_{\alpha}J_{\alpha}\sum_{\alpha_{m}\in\alpha}\prod_{j\in\alpha_{m}}\sigma^{(j)}
\]
We attempt to construct a Hamiltonian as a power series expansion in the mechanical Ising psuedospins of the system. We partition the cones into differing clusters (unique layouts of cones) $\alpha$ with energies $\pm J_{\alpha}$ dependent on the parity of the cluster's cone states. Then for each cluster $\alpha_{m}$ we assign the cones states $\sigma^{(j)}=\pm1$, depending on their orientation. Our expansion includes only clusters with even numbers of cones: odd-numbered clusters are absent due to inversion symmetry. After the clusters are defined, the terms to the right of $J_{\alpha}$ become a correlation matrix:
\[
\Pi_{\alpha k}=\sum_{\alpha_{m}\equiv\alpha}\prod_{j\in\alpha_{m}}\sigma_{k}^{(j)}
\]
where $k$ counts different shapes, and the unknown cluster coefficients $J_{\alpha}$ are found as a solution to a linear system:
\[
E_{k}=\sum_{\alpha}J_{\alpha}\Pi_{\alpha k}
\]
which is done by least squares minimization, using a singular value decomposition pseudoinverse. 

As our initial model attempt produced results whose variations in energy, even for large cluster sizes, are very large, we made another attempt while ignoring the internal symmetries of our system, modeling each cluster as distinct. As an example, this means that every possible group of four cones is modeled as a unique cluster type with its own coefficient. Even then, this does not produce results with any close match to the data until we include six-cone clusters, and high accuracy with eight-cone clusters, at which point we have nearly as many coefficients as unique states in the system. Presumably the reason why the cluster expansion fails is the presence of non-Ising states that fall outside of its purview, combined with a high degree of longer-ranged (and collective) elastic interaction due to the slow power-law falloff of elastic interactions with distance. There is an overall antiferromagnetic trend, as descibed in the main text, but it is swamped by effectively stochastic variations when one looks beyond the main trend.

\begin{table}
\begin{centering}
\begin{tabular}{|c|c|c|c|c|c|}
\hline 
cluster types present & 2(pairwise) & 2,4 & 2,4,6 & 2-8 & 2-10\tabularnewline
\hline 
\hline 
total energy error (eV) & 29997 & 11940 & 3351 & 589 & 440\tabularnewline
\hline 
median energy error (eV) & 1.76 & 1.09 & 0.58 & 0.16 & 0.0007\tabularnewline
\hline 
number of clusters & 22 & 157 & 397 & 532 & 553\tabularnewline
\hline 
\end{tabular}
\par\end{centering}
\caption{Results of cluster model fitting. The top row shows the total error in energy for fits with varying number of cluster types. A small percentage of data points were excluded from the fit due to errors in cone assignment. The middle row shows the median absolute error in energy for each state produced by the fit, and the bottom shows the total number of clusters used for each fit. A very large number of clusters is needed to obtain an accurate energetic description, suggesting a hidden variable related to non-cone configurations.}
\end{table}

\pagebreak

\begin{figure}[t] 
\includegraphics{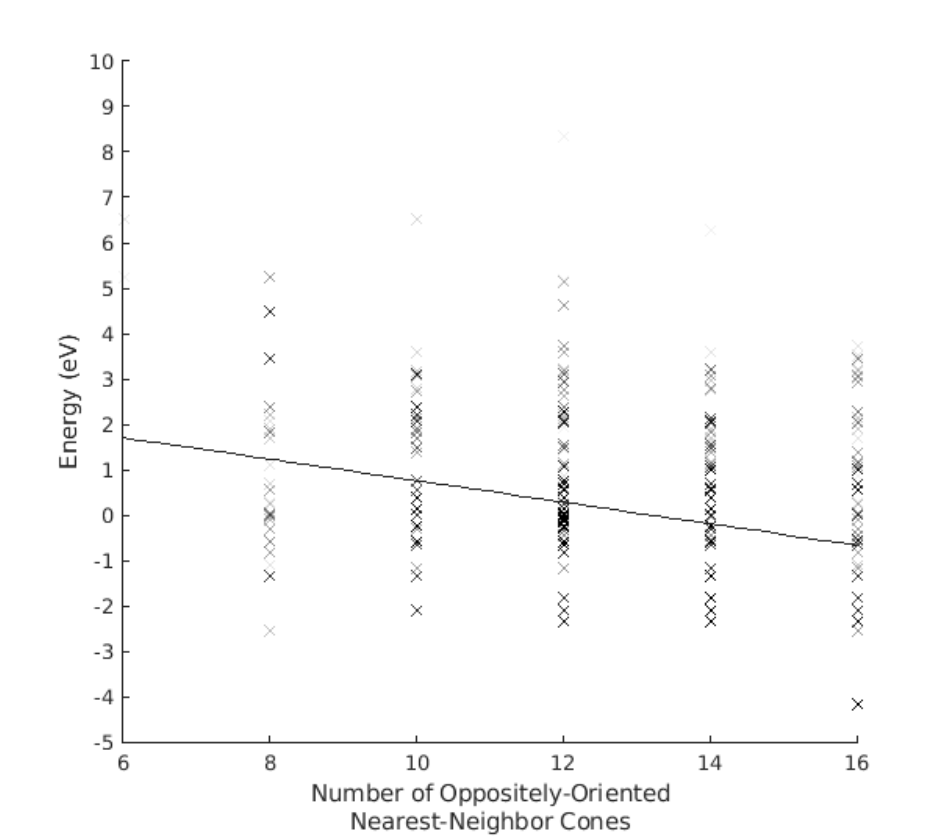}
\caption{The energy for a 1-hexagon-spacing version of the periodic cell studied in the main paper plotted against the number of oppositely-oriented nearest-neighbor cones: this is a 1-hexagon version of Figure 3 of the main text. A linear fit produces $J \sim -0.12$. Although this fit shows significantly more scatter than the six-hexagon-separation case shown in the main paper, the effective antiferromagentic J does scale roughly linearly with the separation between cones (i.e. the effective size of the cones). The Ising states of the most favorable shapes differ from those of the six-hexagon separation case, and certain Ising configurations at the smaller spacing are not metastable (i.e. relax into other configurations). The translucency of the points helps to show clustering around certain Ising states (more so than in the six-hexagon case).}
\end{figure}

\begin{figure}[t] 
\includegraphics[width=6.5in]{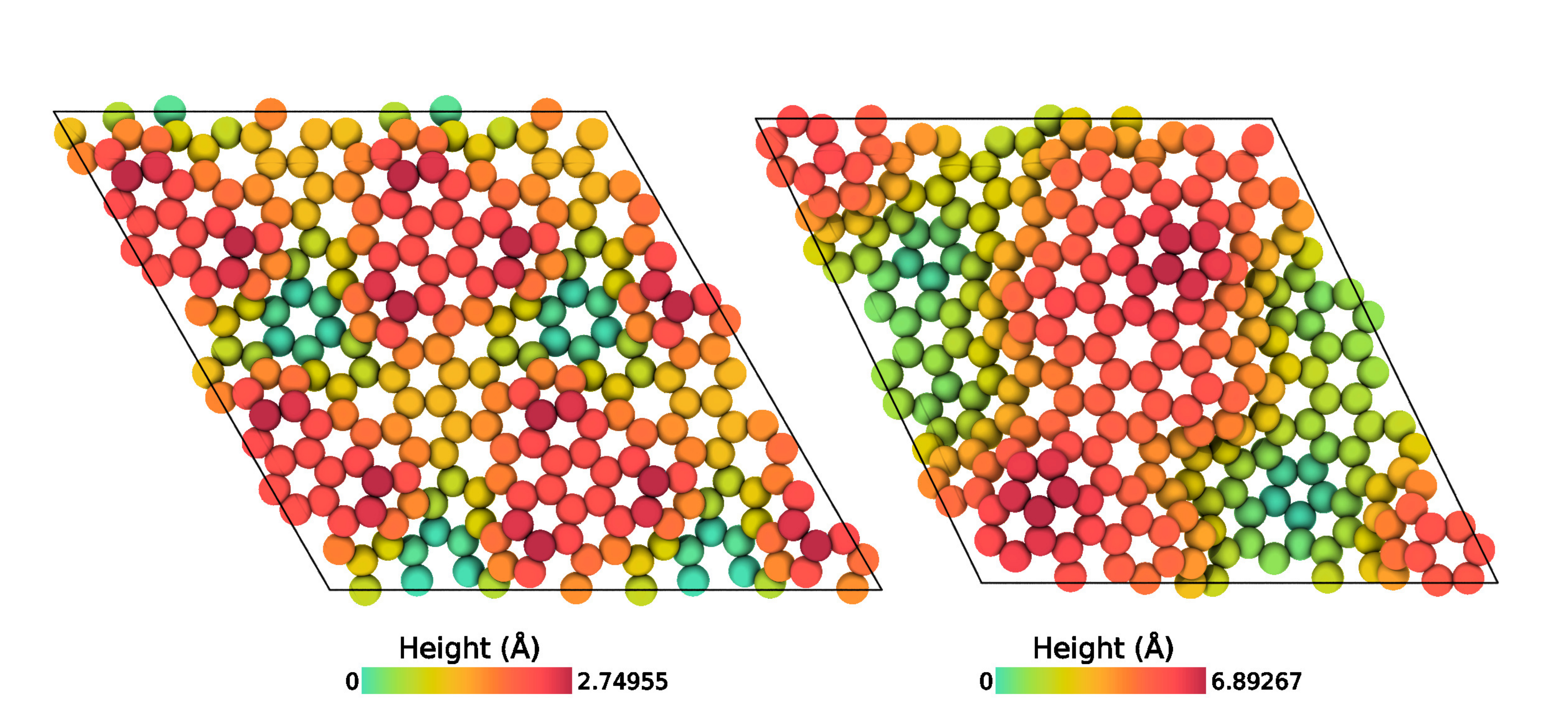}
\caption{Two images of a periodic-boundary 1-hexagon-separation sheet, before (left) and after (right) simulation at 300 K. The left state, which is stable at room temperature for a six-hexagon separation between disclinations, rapidly converts to a new metastable Ising configuration at this closer spacing. At one-hexagon separation, there are fewer metastable Ising states and easier interconversion between these states.}
\end{figure}

\begin{figure}[t] 
\includegraphics[width=6.5in]{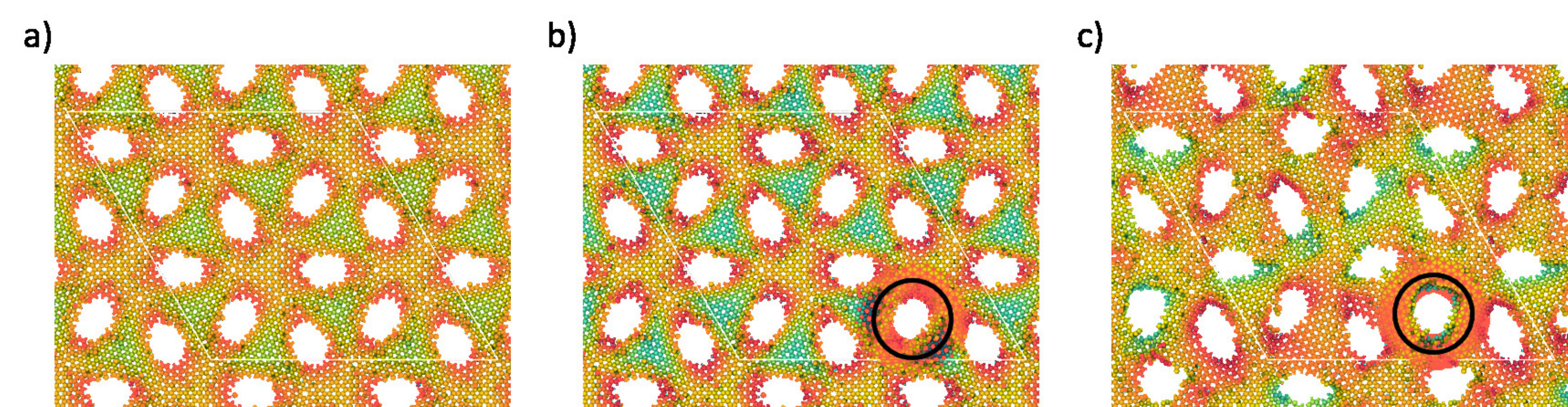}
\caption{NVT simulations of an oxidized sheet with cone apices etched away, in periodic boundary conditions. Panel (a) shows the sheet pre-relaxation, with apices removed and the new boundaries terminated with hydrogen. Panel (b) shows the same sheet post-relaxation. Panel (c) shows this sheet after 1 ns at 800 K: the black circles highlight a cone that spontaneously inverted under thermal excitation. The same sheet without etching does not invert under these conditions: etching the apices greatly reduces the kinetic barrier against inversion at this cone separation.}
\end{figure}

\begin{figure}[t] 
\includegraphics[width=6.5in]{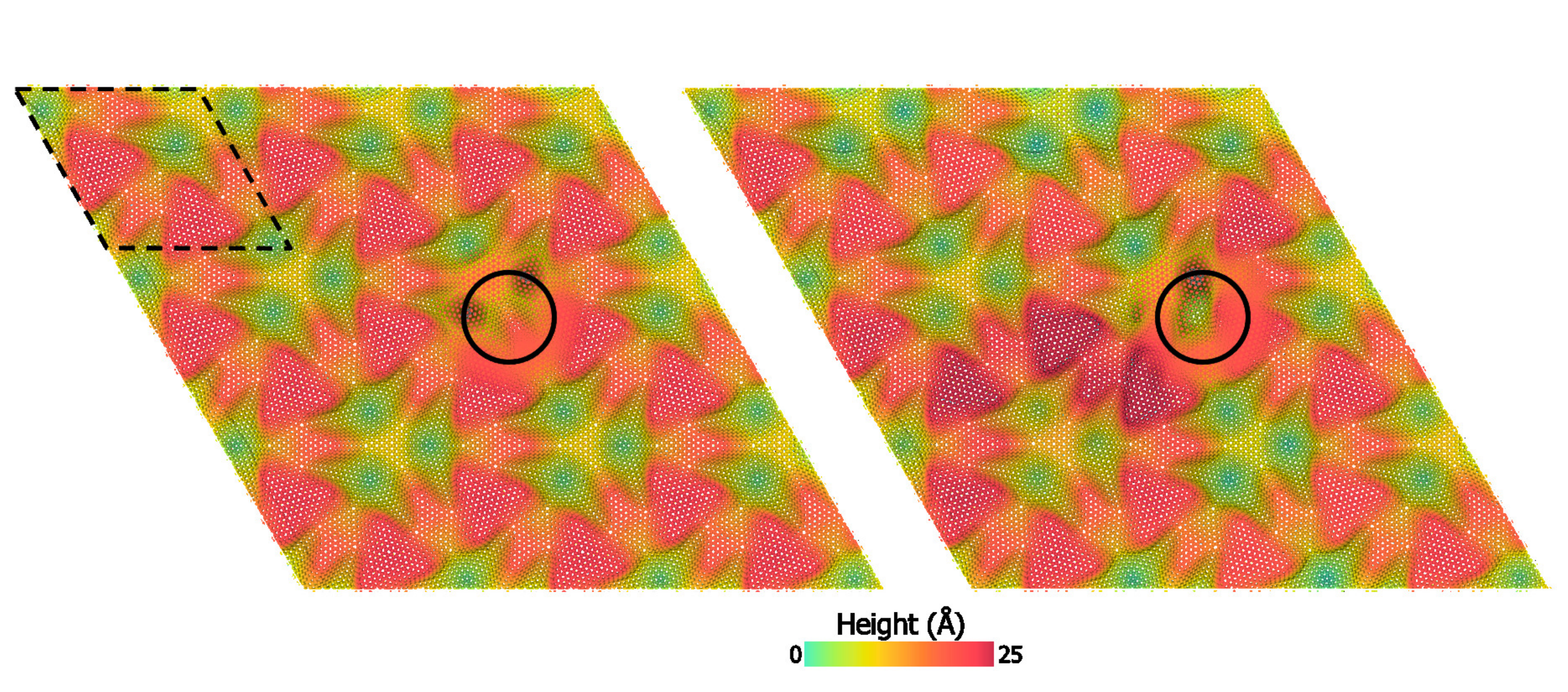}
\caption{A $3\times 3$ supercell of the periodic sheet studied, before and after the inversion of a single cone (circled). The topography of the sheet changes visibly over a region larger than the $1\times 1$ patch studied (denoted with a dashed line), underlining the relatively long range of elastic interactions.}
\end{figure}

\begin{figure}[t] 
\includegraphics[width=6.5in]{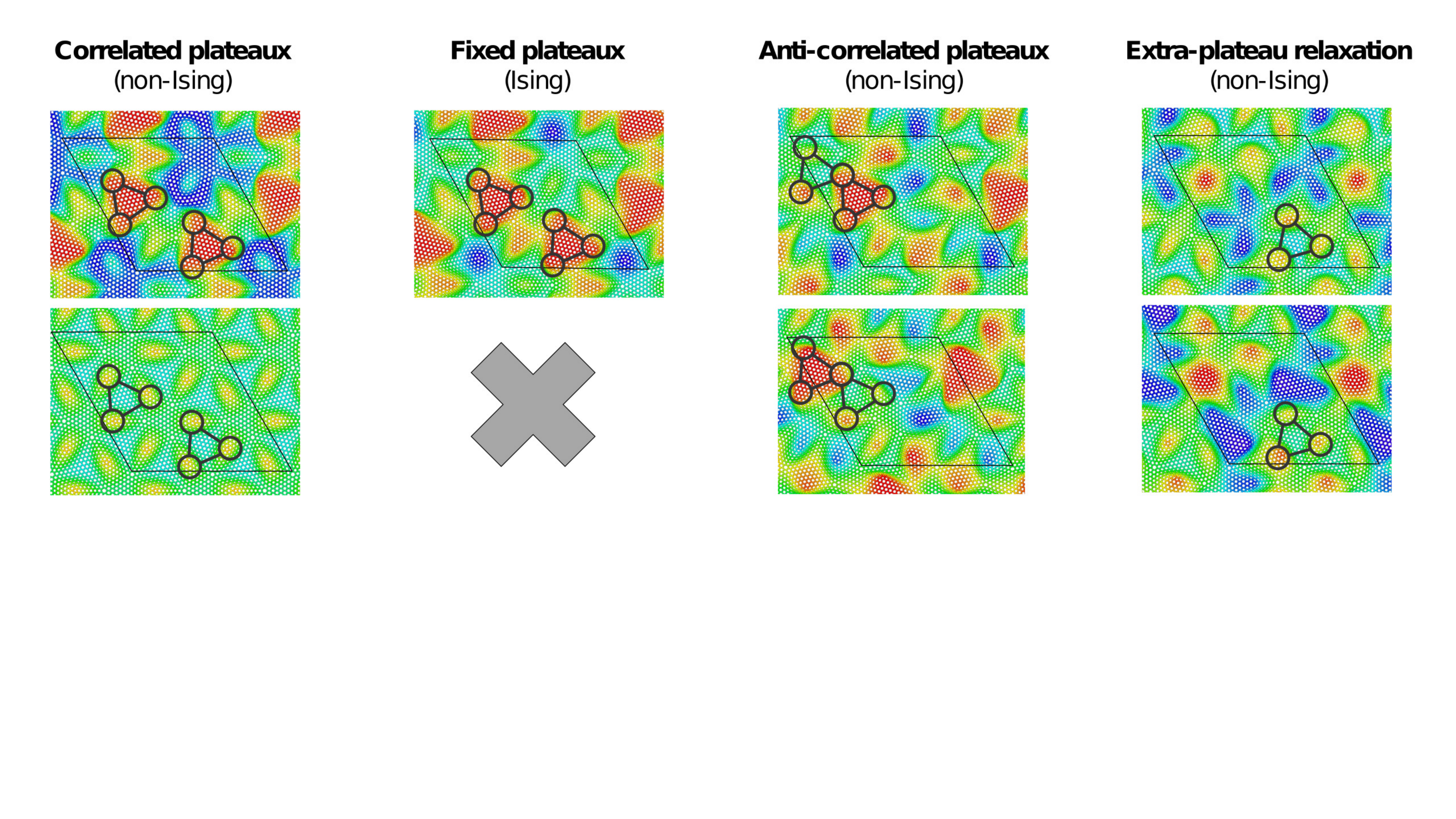}
\caption{Four different cone orientations (top row) each with a candidate ``plateau'' motif bounded by three co-aligned adjacent cones, to explore if manipulation of the plateau region -- while preserving the cone orientations -- can generate non-Ising states. Three of these four cases have non-Ising partners,  but of different characters: the new metastable state is associated either with correlated or anti-correlated shifts of the plateau height or with deformations around but not on the plateau. The second case does not have an Ising partner, although it does have the candidate motif. All shapes shown are metastable under a combination of local relaxation and molecular dynamics at 300 K (similar to the method used in the main text). This diversity of behaviors reflects the absence of a simple rule for the presence or absence of non-Ising states.}
\end{figure}

\appendix

\bibliographystyle{apsrev4-2}
\bibliography{supplement}